\pdfoutput=1
\documentclass[twocolumn]{aastex701}

\usepackage{amsmath}
\usepackage{comment}

\usepackage{bm}
\renewcommand{\vec}{\bm}
\newcommand{\uvec}[1]{\hat{\bm{#1}}}
\newcommand{\mat}{\mathsf}

\newcommand{\gagg}{g_{a\gamma\gamma}}
\newcommand{\DM}{{\rm DM}}
\newcommand{\dblprime}{{\prime\prime}}
\newcommand{\HeI}{He\,\textsc{i}}
\defcitealias{SPHERExPublicProducts}{SPHEREx Science Team 2021}

\shorttitle{Axions in SPHEREx}
\shortauthors{Hainje, Pinetti, \& Hogg}

\usepackage{etoolbox}
\makeatletter
\def\countcols#1{\ifcat#1c%
  \global\advance\colcount by 1\relax\fi
  \def\lookforD{#1}%
  \ifx\lookforD\xD
    \global\advance\howmanyDs by 1
    \global\advance\colcount by1\relax
  \fi
  \futurelet\next\lookatnext}
\patchcmd{\@array@array@new}{\catchpreamble{#2\relax} }{\catchpreamble{#2\relax}}
  {}{\PackageWarningNoLine{aastex-fix}{patch 1 of \string\@array@array@new\space failed}}
\patchcmd{\@array@array@new}{\tabfullpreamble{#2\relax} }{\tabfullpreamble{#2\relax}}
  {}{\PackageWarningNoLine{aastex-fix}{patch 2 of \string\@array@array@new\space failed}}
\patchcmd{\@array@array@new}
  {{\colcount=0 \def\two{#2} \ifx\matrixpreamble\two\else\expandafter\countcols\catchpreamble\global\totalcolumns=\colcount\fi} }
  {{\colcount=0 \def\two{#2}\ifx\matrixpreamble\two\else\expandafter\countcols\catchpreamble\global\totalcolumns=\colcount\fi}}
  {}{\PackageWarningNoLine{aastex-fix}{patch 3 of \string\@array@array@new\space failed}}
\makeatother

\begin{document}

\title{First constraints on axion-like particles with SPHEREx data}

\author[orcid=0000-0002-2506-2502,gname='Connor',sname='Hainje']{Connor~Hainje}
\affiliation{Center for Cosmology and Particle Physics, Department of Physics, New York University, 726~Broadway, New York, NY 10003, USA}
\email[show]{connor.hainje@nyu.edu}
\correspondingauthor{Connor~Hainje}

\author[orcid=0000-0001-7070-0094,gname='Elena',sname='Pinetti']{Elena~Pinetti}
\affiliation{Center for Computational Astrophysics, Flatiron Institute, 162 Fifth Ave, New York, NY 10010, USA}
\email[show]{\newline\hspace*{-1.2em}\hphantom{Email: }epinetti@flatironinstitute.org}

\author[orcid=0000-0003-2866-9403,gname=David,sname=Hogg]{David~W.~Hogg}
\affiliation{Center for Cosmology and Particle Physics, Department of Physics, New York University, 726~Broadway, New York, NY 10003, USA}
\affiliation{Center for Computational Astrophysics, Flatiron Institute, 162 Fifth Ave, New York, NY 10010, USA}
\affiliation{Max Planck Institute for Astronomy, K\"onigstuhl 17, Heidelberg 69117, Germany}
\email[show]{\newline\hspace*{-1.2em}\hphantom{Email: }david.hogg@nyu.edu}

\begin{abstract}
We present the first search for axion-like particle dark matter using SPHEREx data.
We perform a conservative analysis on the first 16.5 months of SPHEREx data and derive new limits on the mass range $0.496 < m_a < 3.32 \, {\rm eV}$ that are competitive with the literature.
Our tightest limit on the coupling relative to those previously published is $\gagg < 2.10 \times 10^{-11} \, {\rm GeV}^{-1}$ at $m_a = 0.606 \, {\rm eV}$, with a median limit of $\gagg < 4.16 \times 10^{-11} \, {\rm GeV}^{-1}$.
Cast as model-independent limits on the lifetime of the dark matter particle, the median limit is $\tau > 6.5 \times 10^{25} \, {\rm s}$.
We are also able to detect several expected emission features at high significance, such as atmospheric \HeI{} and the 3.3~$\mu$m PAH feature.
We calculate the Cram\'er--Rao bound and compare to sensitivity forecasts, finding that our limits could be improved by more than an order of magnitude with solely methodological improvements.
\end{abstract}

\section{Introduction}
\label{sec:intro}
Two of the biggest open problems in fundamental physics are the dark matter (DM) problem and the strong CP problem of quantum chromodynamics (QCD). Evidence for the former comes from numerous probes over different physical scales and timescales (see, e.g., \citealp{Cirelli:2024ssz} for a recent review). Despite the overwhelming evidence for its existence, no experiments can yet irrefutably claim any nongravitational signal from DM particles.  
Regarding the strong CP problem, the community has spent several decades trying to explain why we do not observe a neutron electric dipole moment, despite it being allowed by the QCD Lagrangian.
A possible explanation is that a new field---the axion---dynamically relaxes the effective CP-violating angle of QCD toward zero, suppressing the neutron electric dipole moment (see, e.g., \citealp{DiLuzio:2020wdo} for a review).
This QCD axion is also a promising DM candidate, potentially providing a common explanation for both puzzles.
More generally, axion-like particles (ALPs) arise in many extensions of the Standard Model and can also constitute DM, although ALP masses and couplings need not obey the QCD axion relation and ALPs do not necessarily solve the strong CP problem (see, e.g.,~\citealp{OHare:2024nmr} for a recent review).

Blank-sky observations offer a promising avenue for DM searches by probing regions with limited contamination from bright astrophysical sources.
These fields contain DM along the line of sight, whose decay into photons could produce a narrow spectral line. Large samples of archival blank-sky observations can be combined to increase sensitivity to such faint signals. 

The Spectro-Photometer for the History of the Universe, Epoch of Reionization, and Ices Explorer (SPHEREx) is a NASA mission that is currently conducting the first all-sky near-infrared (0.75--5~$\mu$m) spectral survey \citep{Bock+2026}.
Over its 25-month mission, it will cover the full sky four times, as well as two $\sim 100$ square degree Deep Fields near the north and south ecliptic poles at up to $\sim 100$ times more coverage.
Because the decay of axion-like particles produces a narrow emission line with wavelength inversely related to their mass, the SPHEREx wavelength range corresponds to axion masses 0.496--3.32 eV.
This range has overlap with relatively few other instruments:
    MUSE covers 2.66--5.27 eV \citep{Todarello:2023hdk, Regis:2020fhw},
    WINERED covers 1.83--2.73 eV \citep{Yin:2024lla}, and
    JWST NIRSpec covers 0.45--4.1 eV \citep{Pinetti2025, JanishPinetti2025,Saha:2025any,Gemmell+2026, Roy:2023omw}.
Of these, NIRSpec is the most analogous to SPHEREx:
NIRSpec has much higher resolution,
but SPHEREx has vastly greater sky coverage and blank-sky data volume.
The SPHEREx Deep Fields in particular feature a substantial amount of data with relatively low foregrounds from Galactic or zodiacal light emission and strong forecast sensitivity \citep{Regis+2025}.

In this work, we perform a conservative search for anomalous emission.
At high significance, we are able to detect a few known and unmodeled spectral features, such as atmospheric \HeI{} and the 3.3~$\mu$m polycyclic aromatic hydrocarbon (PAH) feature.
We do not find any unknown spectral lines at high significance, and we set 95\% power-constrained upper limits on the coupling $\gagg$ across the mass range.
These limits are competitive with those in the literature.
We additionally compute the Cram\'er--Rao bound and the mission-complete forecast, finding evidence that, with better methods, there is substantial room for improvement.
The code for this project is made open source under an MIT license on GitHub at \url{https://github.com/cmhainje/spax}.

\section{Data}
\label{sec:data}

SPHEREx observes with six detectors, each consisting of a $2040 \times 2040$ array of $6.15^\dblprime$ square pixels.
Each detector measures a different wavelength band and is fitted with a Linear Variable Filter (LVF), so each pixel measures a narrow spectral bandpass with central wavelength varying continuously over the detector.
The wavelength coverage of these six bands is given in Table~\ref{tab:wavelength}.
The coverage is nominally divided into 102 spectral channels, 17 per band, which are approximately spaced by the local bandwidth.
Note that we do \textit{not} bin the data by spectral channel; we preserve the full, continuous wavelength information of each pixel.

\begin{table}[t]
    \centering
    \begin{tabular}{ccc}
        \hline
        Band & Range [$\mu$m] & Resolution $R$ \\
        \hline
        D1 & 0.744--1.116 & 39 \\
        D2 & 1.099--1.651 & 41 \\
        D3 & 1.636--2.421 & 41 \\
        D4 & 2.423--3.822 & 35 \\
        D5 & 3.809--4.420 & 112 \\
        D6 & 4.412--5.002 & 128 \\
        \hline
    \end{tabular}
    \caption{Wavelength coverage of the SPHEREx instrument \citep{ExpSupp}.}
    \label{tab:wavelength}
\end{table}

As part of the SPHEREx ``Quick Release'' data products, all exposures are calibrated and released within 60 days of observation \citep{ExpSupp}.
These data are delivered as ``spectral images'': one per detector per exposure, consisting of the calibrated intensity measurement, variance, status flags, zodiacal light estimate, and astrometric solution for each pixel in the detector.
Because of the LVF design, spectral coverage is built up gradually by revisiting each point on the sky and placing it in different locations in the field of view until its brightness has been measured in each of the 102 spectral channels.
This creates a challenge when dealing with time-varying signals and contaminants (such as atmospheric emission or zodiacal light).
Because the DM halo and our sightline through it do not vary significantly, we assume that the sky is static, and temporal variations in the data are an additional source of noise.

\begin{figure*}
    \centering
    \includegraphics[width=\linewidth]{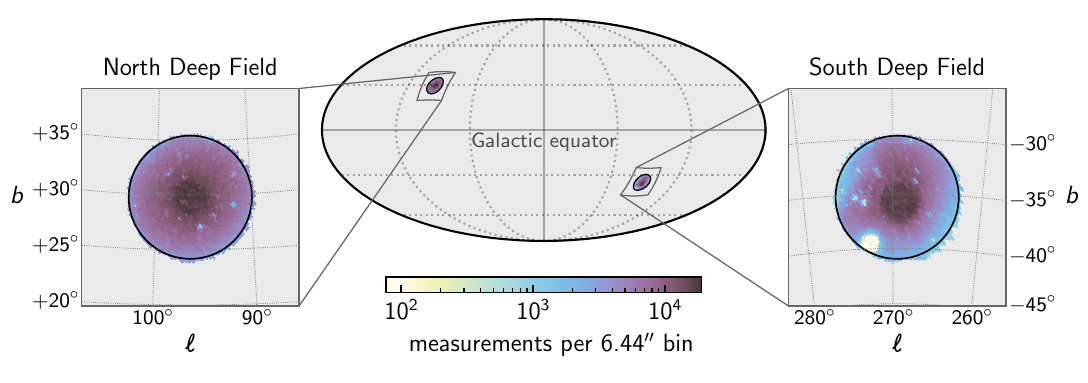}
    \caption{Sky distribution of the data we used.
    This includes every pixel measurement in the Deep Fields from QR2 or QR3 (up to week 36 of 2026) that passes our positivity and status-flag cuts.
    The distribution is visualized as the number of pixel measurements per $N_{\rm side} = 2^{15}$ HEALPix pixel.
    Full spectral coverage in a HEALPix pixel corresponds to 102 pixel measurements.
    In black, $5.64^\circ$ circles are drawn; for our analysis, we use all measurements whose parent partition (roughly, its $N_{\rm side} = 2^8$ HEALPix index) intersects the circle.
    }
    \label{fig:coverage}
\end{figure*}

We use all Quick Release version 2 (QR2) and 3 (QR3) spectral images, up to exposures taken in week 36 of 2026 (representing the first 16.5 months).
Between QR2 and QR3, the processing pipeline was substantially updated, with upgrades to the calibrations and the zodiacal light estimate.
Calibration data products (flux calibration, absolute gain matrix, dark current map, zodiacal light scaling) enable us to recalibrate QR2 data to match QR3.
For a QR2 measurement from detector pixel $j$ with intensity $I$, variance $V$, and zodiacal light $Z$, the QR3-recalibrated measurement is given by
\begin{align}
    I_{\rm QR3} &= c_j \, I_{\rm QR2} + A_{{\rm QR3},j} \, (D_{{\rm QR2},j} - D_{{\rm QR3},j}), \\
    V_{\rm QR3} &= c_j^2 \, V_{\rm QR2}, \\
    Z_{\rm QR3} &= \alpha(\lambda_j) \, Z_{\rm QR2},
\end{align}
where
$c$ is the flux correction data product,
$A$ is the absolute gain matrix,
$D$ is the dark current map,
and $\alpha$ is the zodiacal light scaling, linearly interpolated to the central wavelength $\lambda_j$ of pixel $j$.

The spectral images are then preprocessed into a pixel-level database called the \texttt{talltable} (an article describing the details of this is in preparation).
To summarize, we open every spectral image and ``flatten'' it into a set of tuples containing \{right ascension, declination, central wavelength, bandwidth, intensity, variance, status flags, zodiacal light model\}.
Doing this for all images results in $6.3 \times 10^{12}$ such tuples, which we sort by a very high-resolution nested HEALPix index \citep{HEALPix} and store in chunks called ``partitions''.
The partitions are defined by coarser HEALPix indices at $N_{\rm side} = 2^6$--$2^9$, depending on the local density of measurements.

For this analysis, we choose to use data from both the North Deep Field (NDF), centered on the north ecliptic pole, and the South Deep Field (SDF), centered at an $8^\circ$ offset from the south ecliptic pole to avoid the LMC \citep{Bock+2026}.
Because of their position, the Deep Fields are expected to have lower foregrounds (from zodiacal light or Galactic emission) and fewer sources compared with the rest of the sky.
With reduced nuisances and substantial coverage, the Deep Fields are forecast to be highly sensitive to axion emission \citep{Regis+2025}.\footnote{%
\citet{Regis+2025} forecast that the sensitivity of the LMC could rival that from the Milky Way, in large part because of the overlap between the LMC and the SDF under a previous survey plan.
However, with the offset of the SDF center from the south ecliptic pole, there is practically no overlap, and the SDF can be easily included in a Milky Way analysis.}
We select the Deep Field data by using all partitions that come within $5.64^\circ$ of the NDF or SDF centers, totaling 213 square degrees.\footnote{Instructions for downloading these partitions can be found at \url{https://github.com/cmhainje/spax/blob/main/talltable_paths.txt}.}

We filter the resulting set of pixel measurements by requiring positive, finite values for the intensity and variance and no status flags set.
Importantly, one status flag, the ``known source'' flag, is set for any measurement that is expected to receive significant flux from a known source in the SPHEREx Reference Catalog, with the masking radius of each source set according to its estimated magnitude.\footnote{%
\citet{ExpSupp} note that, because the size of the source mask is determined by the magnitude, errors in the magnitude can result in source masks being too small or too large.}
About 50\% of all measurements in these fields are marked with the known-source flag.
Filtering out these measurements gives us a blank-sky sample, allowing us to avoid the modeling of bright astrophysical sources.

\begin{figure*}
    \centering
    \includegraphics[width=\linewidth]{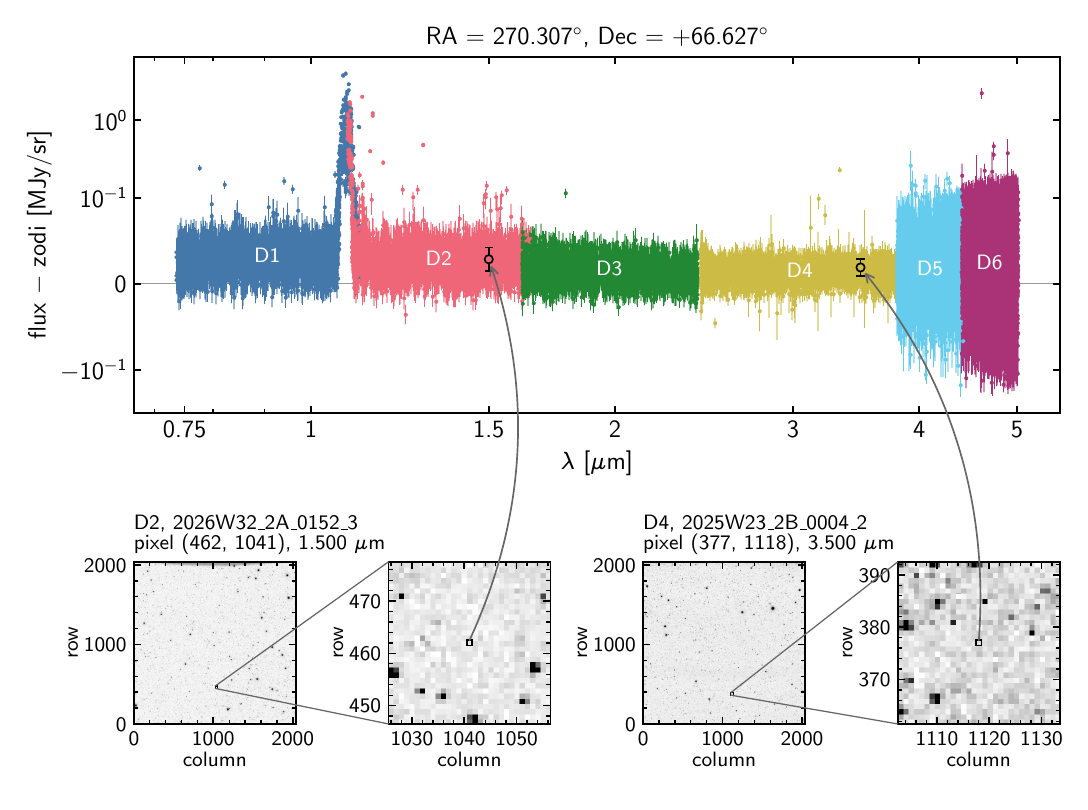}
    \caption{A sample spectrum, located $8^\prime$ from the north ecliptic pole, chosen for being relatively dark (low median of zodiacal-light-subtracted intensity, and low number of measurements flagged as ``known source'').
    The $N_{\rm side} = 2^{15}$ bin centered at this location consists of 28,848 measurements after our positivity and status-flag cuts.
    Detector bands are marked by different colors; bands D5 and D6 have higher resolution and therefore smaller range.
    The massive bump at 1.083 $\mu$m is an atmospheric \HeI{} line.
    Two pixel measurements are selected arbitrarily at 1.5 and 3.5 $\mu$m and given a black outline; the source spectral images and a zoomed-in cutout of these measurements are shown below.
    }
    \label{fig:spectrum}
\end{figure*}

Finally, we bin the measurements into HEALPix pixels with $N_{\rm side} = 2^{15}$, which are roughly $6.44^\dblprime$ in size, very similar to the SPHEREx pixel size.
The distribution on the sky of the resulting $4.70 \times 10^{11}$ pixel measurements is shown in Figure~\ref{fig:coverage}.
Each bin is then treated as a single ``spectrum'' for downstream processing; a sample is shown in Figure~\ref{fig:spectrum}.

In addition to the pixel measurements, we use the spectral calibration data product (version \texttt{v5-2026-191}), the spectral channels data product (version \texttt{v2-2026-229}), and the response curves published in \citet{SpecFiles} for the most precise wavelength solutions.
Specifically, we use the 68-subchannel central-pixel response curves.
As recommended, when working with a given pixel $j$, we look up the nearest central-pixel response curve and shift it so that its central wavelength matches that of pixel $j$.

\section{Method}
\label{sec:method}

Much of the method used is taken from \citet{JanishPinetti2025} and \citet{Pinetti2025},
with modifications to adapt to the specific properties of the SPHEREx instrument.
We adopt natural units with $\hbar = c = 1$.

\subsection{Axion emission intensity}

We suppose the dark matter (DM) is entirely composed of axions or ALPs of mass $m_a$.
They decay isotropically to two photons with rate
\begin{equation}
    \Gamma = \frac{\gagg^2 \, m_a^3}{64 \pi} .
\end{equation}
The photons each have energy $m_a / 2$, so the luminosity per DM particle is $\Gamma \, m_a$.

Suppose the velocity distribution of the DM particles is an isotropic Maxwellian with width $\sigma_v$.
In the limit that the DM particles are nonrelativistic, the photons will have observed frequencies that are Gaussian-distributed with mean $\nu_a = m_a / 4\pi$ and standard deviation $\sigma_v \, m_a / 4\pi$.
We denote the probability density of observed frequency $\nu$ as $p(\nu)$.

The total specific intensity of axion emission incident on our detector from line of sight $\uvec\alpha$ is given by
\begin{equation}
    I_\nu^{\rm inc} (\uvec\alpha) = 
    \frac{\Gamma}{4 \pi} \, p(\nu) \,
    \int_{0}^{\infty} ds \, \rho_{\DM} (r [s, \uvec\alpha]) ,
\end{equation}
where the line-of-sight integral over the DM density distribution $\rho_{\DM}(r)$ ($r$ the spherical radius) is denoted $D(\uvec\alpha)$ and called the ``$D$-factor''.

Each pixel in the SPHEREx instrument has a unitless, peak-normalized spectral response curve $T_j (\lambda)$ ($j$ being the index of a pixel).
The reported intensity measured by pixel $j$ when subject to specific intensity $I_\nu$ is the wavelength-space bandpass average of the incident intensity \citep[Eq.~22]{ExpSupp}.
Therefore, the intensity of axion emission reported by pixel $j$ when pointing along $\uvec\alpha$ is
\begin{equation}
    I_\nu^{\DM} (\uvec\alpha; j) =
    \frac{\Gamma \, D (\uvec\alpha)}{4 \pi} \,
    \frac{1}{\Delta \lambda_j} \,
    \int d\lambda \, T_j(\lambda) \, p(\nu) \big|_{\nu=1/\lambda} ,
\end{equation}
where $\Delta \lambda_j = \int d\lambda \, T_j(\lambda)$ is the bandwidth of pixel $j$.

In this work, we only consider emission from the Milky Way DM halo.
For the halo density profile, we use an NFW profile with scale radius $r_s = 14.46 \, {\rm kpc}$ and scale density $\rho_s = 0.566 \, {\rm GeV/cm}^3$ \citep{Cirelli:2024ssz}, as used in the forecasts of \citet{Regis+2025}.
For the velocity dispersion, we use $\sigma_v = 160 \, {\rm km/s}$ \citep{Evans+2018}.

\subsection{Sensitivity forecasts}

Following~\citet{Cowan+2011-likelihood}, we define the Asimov test statistic as 
\begin{equation}
    \Delta \chi_A^2 = \sum_i \frac{I_\nu^{\DM} (\uvec{\alpha}_i; j_i)^2}{\sigma_i^2},
\end{equation}
where $i$ indexes any single pixel measurement,
$\uvec{\alpha}_i$ is the position on the sky of the measurement (snapped to the nearest $6.44^\dblprime$ HEALPix pixel center for compatibility with our later analysis),
$j_i$ is the index of the detector pixel that performed the measurement,
and $\sigma_i$ is the reported uncertainty.
The Asimov expected sensitivity, i.e., the median expected one-sided 95\% upper limit on $g_{a \gamma \gamma}$, is the coupling at which $\Delta \chi_A^2 = 2.71$ for a given mass $m_a$.

Our sensitivity is computed using the masked and filtered data.
\citet{ExpSupp} note that the observed sensitivity of the instrument is ``as-good-as or better than the original projected performance estimates'' in all channels except the \HeI{}-contaminated 1.083 $\mu$m channel.
However, because the known-source mask removes about 50\% of measurements, we find slightly worse surface-brightness sensitivity than was predicted pre-launch \citepalias{SPHERExPublicProducts}.

Because our data represent only the first 16.435 months, we forecast the mission-complete sensitivity forward to 25 months via a linear scaling:
\begin{equation}
    \Delta \chi_{A, 25}^2 = \frac{25}{16.435} \, \sum_i \frac{I_\nu^{\DM} (\uvec{\alpha}_i; j_i)^2}{\sigma_i^2}.
\end{equation}
Because $I_\nu^{\DM} \propto \gagg^2$, this change amounts to scaling the upper limits on $\gagg$ by $(25/16.435)^{-1/4} = 0.900$.
These forecasts are consistent with the corrected sensitivity forecast of \citet{Regis+2025}.\footnote{The projections in the journal version of \citet{Regis+2025} are weaker 
because their calculation compares the line-integrated intensity $I$ to the per-channel $\nu I_\nu$ sensitivity. 
Accounting for the channel width increases the expected signal by a factor of the resolution $R$, improving the forecast sensitivity to $\gagg$ by $\sqrt{R} \approx 6$--$11$ across the SPHEREx bands.
Because of our masking, our forecast does not realize this full factor and is instead 5--8 times stronger than their journal-version projections. After detailed discussions with the authors, their revised arXiv version (v3) is now consistent with our predictions.} 

\subsection{Statistical analysis}

As mentioned previously, we spatially bin the pixel measurements into $N_{\rm side} = 2^{15}$ HEALPix pixels and treat each bin as a single ``spectrum''.
Additionally, to avoid issues in modeling that may arise at detector boundaries, we split each such bin into groups by detector.
Let $b$ index a (HEALPix pixel, detector) pair.

After preprocessing, this bin consists of a set of pixel measurement tuples containing the specific intensity $I_{\nu, i}$, zodiacal light estimate $Z_i$, uncertainty $\sigma_i$, and the measuring detector pixel index $j_i$.
Detector pixel $j_i$ has an associated spectral response curve $T_{j_i}(\lambda)$, central wavelength $\lambda_{j_i}$, and bandwidth $\Delta \lambda_{j_i}$.

Following \citet{Cirelli+2021}, \citet{Cowan+2011-likelihood}, \citet{Roach+2023}, and \citet{Pinetti2025}, we define the goodness of fit:
\begin{equation}
  \chi^2(g_{a\gamma\gamma},\boldsymbol{\theta}\,|\,m_a)
  = \sum_b \sum_{i\in b} \frac{r_i^2}{\sigma_i^2},
  \label{eq:chi2}
\end{equation}
where the residual of measurement $i$ is
\begin{equation}
  r_i \equiv I_\nu^{\rm DM}(\hat{\boldsymbol{\alpha}}_b; j_i)
  + \mathcal{B}(\lambda_{j_i};\boldsymbol{\theta}) + Z_i - I_{\nu,i} ,
  \label{eq:residual}
\end{equation}
where $\mathcal{B}$ is a continuum model with parameters $\vec\theta$.
Here we make a conservative choice: we use an independent continuum $\mathcal{B}$ for every bin $b$.
This affords the continuum great flexibility to explain the data.
Additionally, we use a smooth functional form (a fourth-order polynomial), fit to a small window ($\pm 5 \, \lambda_a / R$) of the spectrum centered on the axion line.
This limits the flexibility of the continuum somewhat: it cannot fit away any sharp features, which means it also cannot explain known features (such as the \HeI{} line).
The window size reflects a trade-off: it must be wide enough that the continuum is constrained by data outside the line, yet narrow enough for a low-order polynomial to describe the continuum.
Because the intrinsic width of the axion line ($\sigma_v/c \approx 5 \times 10^{-4}$) is much smaller than the instrumental resolution ($1/R \approx 0.008$--$0.03$), the observed line profile is set by the spectral response and has a width of order $\lambda_a/R$, and we therefore choose the window size to span $\sim$10 resolution elements.
We tested on a small patch of sky a sixth-order polynomial, splines with two different knot spacings, and a window size of $\pm 10 \, \lambda_a / R$, none of which had any significant effect on the results.

By choosing $\mathcal{B}$ to be a linear model that is independently fit for each bin, we can analytically profile out the continuum parameters.
The resulting profiled goodness of fit $\chi^2_{\rm p}$ is quadratic in the emission rate $\Gamma$.
The negative logarithm of the profile likelihood is one-half the profiled goodness of fit (up to a constant), so the maximum-likelihood estimator $\hat\Gamma$ and the Cram\'er--Rao bound are simple to derive in terms of the quadratic coefficients.
We require the emission rate to be non-negative, so we define 
$\Delta \chi^2 (\Gamma) = \chi^2_{\rm p} (\Gamma) - \chi^2_{\rm p} ( \max(\hat\Gamma, 0) )$
and set one-sided 95\% upper limits where $\Delta \chi^2 = 2.71$ (i.e., $1.645^2$). Independently, we search for detections using the two-sided 95\% confidence interval on $\hat\Gamma$, which corresponds to  $\Delta \chi^2 = 3.84$ (i.e., $1.96^2$). Its lower limit is positive whenever $\hat\Gamma / \sigma_{\hat\Gamma} > 1.96$, corresponding to a one-sided false-positive rate of 2.5\% under the null hypothesis.

\begin{figure*}[ht!]
    \centering
    \includegraphics[width=\linewidth]{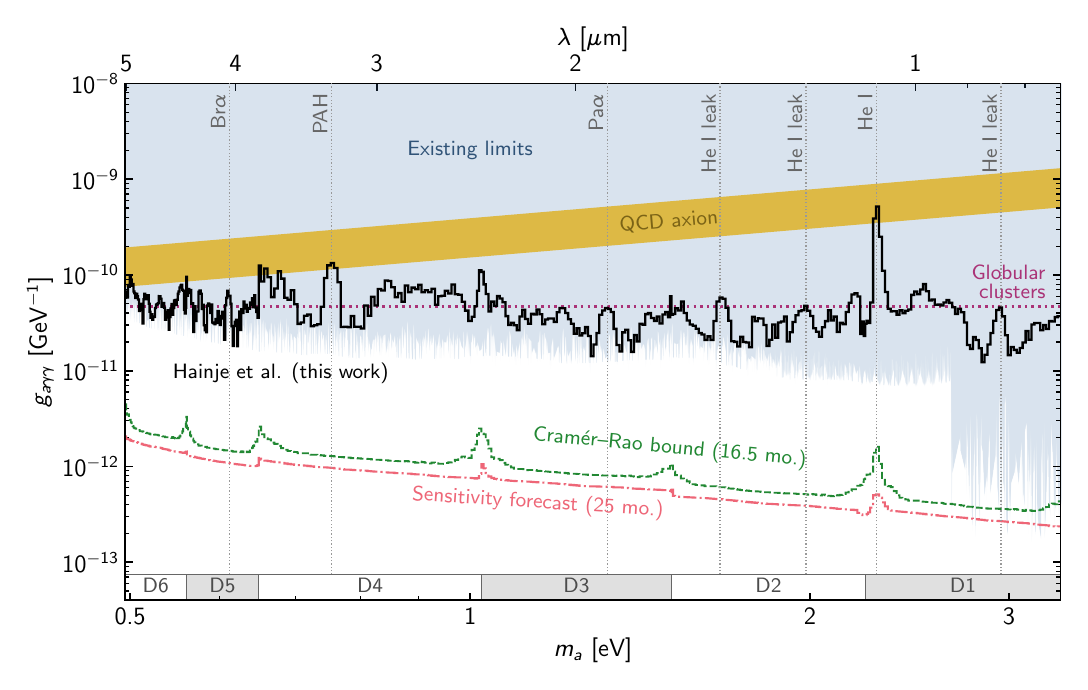}
    \caption{Our results (black), shown as the 95\% power-constrained upper limit on the coupling $\gagg$ as a function of axion mass $m_a$.
    The Cram\'er--Rao bound is shown in a dashed green line.
    The mission-complete sensitivity forecast is shown in a dash-dotted pink line.
    Existing limits, described in the text, are shown in light blue, and the region of parameter space corresponding to the QCD axion is shown in gold.
    The approximate wavelength ranges of the six detector bands are shown along the bottom, and a number of spectral lines are labeled at the top.
    Of note are \HeI{}, three ``\HeI{} leak'' lines (where the out-of-band response is particularly sensitive to the 1.083 $\mu$m \HeI{} line), the H lines Paschen-$\alpha$ (Pa$\alpha$) and Brackett-$\alpha$ (Br$\alpha$), and the polycyclic aromatic hydrocarbon (PAH) 3.3~$\mu$m feature.
    }
    \label{fig:constraint}
\end{figure*}

To be maximally conservative, we power-constrain our upper limits and perform three data-driven error corrections.
The power-constrained limit \citep{Cowan+2011-PCL} is derived using a threshold of 0.1587, which softens exclusions that arise under downward fluctuations larger than $1 \sigma$.
Implementation details are given in Appendix~\ref{appx:pcl}.
The error corrections follow the data-driven error calibration scheme of \citet{Pinetti2025} and are described in detail in Appendix~\ref{appx:error-calibration}.
Briefly, one error correction inflates the uncertainties of individual measurements if they are less than the empirical local scatter,
and the other corrections inflate the uncertainty of the estimator to better normalize the distribution of the $z$-score $\hat\Gamma / \sigma_{\hat\Gamma}$ (since $z \sim \mathcal{N}(0, 1)$ under the null hypothesis).
The second class of corrections is useful for dealing with our smooth continuum model;
because of its smoothness, the residuals have a great deal of structure that does not average down over bins.
This produces excess scatter, and so these corrections inflate the resulting uncertainty on the estimator.

\section{Results}
\label{sec:results}

\begin{table*}[t!]
    \centering
    \begin{tabular}{@{\extracolsep{\fill}}cccl}
        \hline
        $\lambda$ [$\mu$m] & $m_a$ [eV] & $z$-score & Known nearby features \\
        \hline
        0.836--0.851 & 2.91--2.97 & 6.2  & \HeI{} leak / atmospheric (0.84\,$\mu$m) \\
        1.068--1.087 & 2.28--2.32 & 15.8 & \HeI{} 1.083\,$\mu$m \\
        1.122--1.143 & 2.17--2.21 & 6.3  & atmospheric (1.13\,$\mu$m) \\
        1.243        & 1.99       & 2.1  & \HeI{} leak (1.25\,$\mu$m) \\
        1.469--1.496 & 1.66--1.69 & 5.5  & \HeI{} leak (1.49\,$\mu$m) \\
        1.740        & 1.43       & 2.0  & --- \\
        1.867--1.889 & 1.31--1.33 & 3.9  & Pa$\alpha$ 1.875\,$\mu$m \\
        2.531--2.549 & 0.97--0.98 & 2.9  & D3/D4 dichroic \\
        3.238--3.330 & 0.74--0.77 & 17.0 & 3.3\,$\mu$m PAH \\
        3.855        & 0.64       & 2.3  & D4/D5 boundary \\
        4.046--4.064 & 0.61       & 8.2  & Br$\alpha$ 4.051\,$\mu$m \\
        4.291--4.300 & 0.58       & 2.1  & --- \\
        4.464--4.497 & 0.55--0.56 & 3.4  & D5/D6 boundary \\
        4.939--4.949 & 0.50       & 2.4  & red end of band \\
        \hline
    \end{tabular}
    \caption{%
    Wavelength regions for which the lower limit on the 95\% confidence interval is positive ($z > 1.96$ after data-driven error corrections).
    Contiguous test masses are grouped, and the column ``$z$-score'' gives the largest seen score in each region.
    We list nearby features from \citet{ExpSupp} and known astrophysical lines.
    A similar number of regions (16) have $z < -1.96$; see the text.
    }
    \label{tab:detections}
\end{table*}

The 95\% power-constrained upper limit on the coupling $\gagg$ is shown as a function of axion mass $m_a$ in black in Figure~\ref{fig:constraint}.
The upper limits have been computed at 408 test masses, subdividing the nominal spectral channels by a factor of 4.
We additionally show the Cram\'er--Rao bound in a dashed green line, computed under the assumption of independent Gaussian noise with the reported uncertainties of the data (inflated according to the local empirical scatter within each spectrum, Correction I in Appendix~\ref{appx:error-calibration}) after profiling out the continuum model,
and the mission-complete forecast in a dash-dotted pink line, computed only from the observed uncertainties.
The gold band shows the region of parameter space corresponding to the QCD axion, where axions solve the strong CP problem of QCD.
Existing limits are shown in light blue.
The limits comprising this region are from
    the CAST helioscope experiment \citep{CAST:2007jps,CAST:2017uph,CAST:2024eil}
    and line searches with MUSE \citep{Regis:2020fhw,Todarello:2023hdk},
    WINERED \citep{Yin:2024lla},
    and JWST \citep{JanishPinetti2025,Saha:2025any,Pinetti2025,Gemmell+2026}.
We also show a flat upper limit from stellar evolution, based on measurements of globular clusters, in purple \citep{Ayala:2014pea, Dolan:2022kul}.
Most existing limits were drawn from the collection of \citet{AxionLimits}.
We compare the limits as they were published; some studies use different NFW parameters for the Milky Way halo, and some use different $\Delta \chi^2$ thresholds.

In the figure, we additionally label the regions covered by each detector band, as well as a few known and expected astrophysical lines.
There are substantial sensitivity losses at each detector boundary, which are expected as our method treats data from each detector independently, and near boundaries the continuum model is more free to fit away a line.
Additionally, the very bright atmospheric \HeI{} line loosens the upper limit and reduces sensitivity.
There are three wavelengths where the out-of-band response is particularly sensitive to the \HeI{} line; these are labeled as ``leak'' lines.
Some of the brightest expected astrophysical features are the H lines Paschen-$\alpha$ and Brackett-$\alpha$ and the 3.3~$\mu$m PAH feature: these create bump-like features in the upper limit.

These upper limits can be recast as the 95\% power-constrained lower limits on the lifetime $\tau$ of the DM particle, agnostic of the axion model.
These limits are shown in Appendix~\ref{appx:tau};
the median limit is $\tau > 6.5 \times 10^{25} \, {\rm s}$.

We also compute the 95\% confidence interval on the maximum-likelihood estimator $\hat\Gamma$, here omitting the power constraint, to search for ``detections'' (where the lower limit is positive).
We find 14 such regions in wavelength, listed in Table~\ref{tab:detections}.
Importantly, all of the strongest excursions ($z \gtrsim 4$) are explained by nearby known spectral features: \HeI{}, 3.3~$\mu$m PAH, \HeI{} leak lines, Pa$\alpha$, Br$\alpha$, and atmospheric lines as cataloged in \citet{ExpSupp}.
Of the remainder, many are near features of the instrument (detector boundaries), but the detections are not very strong.
The look-elsewhere effect also plays a role: with $\sim 100$ independent spectral channels and a false-positive probability of $2.5\%$, we expect $\sim 2.5$ excesses from Gaussian noise alone.
In addition, compared to the 14 regions identified in the table, there are 16 regions with $z < -1.96$, indicating that excursions with $z > 1.96$ cannot necessarily be identified as emission lines.
The best evidence for having detected emission lines at the points where $z \gtrsim 4$ is their excellent alignment with features that were expected to be strong \textit{a priori}.

\section{Conclusions}
\label{sec:conclusions}

We present the first limits on axion-like particle DM using SPHEREx data.
These limits are very conservative, owing to the use of power-constrained limits, data-driven error calibration, and an independent, smooth continuum model at every point on the sky.
Still, using the first 16.5 months of data, we obtain limits on the coupling $\gagg$ over the mass range $0.496 < m_a < 3.32 \, {\rm eV}$ that are competitive with those seen in other analyses.
The median gap between our limits and the Cram\'er--Rao bound is $39 \times$, and $55 \times$ to the mission-complete sensitivity forecast, indicating that substantially stronger limits are possible.

The upcoming SPHEREx Data Release 1 (DR1) will include reprocessings of all the year 1 data, improving calibrations, bad-pixel flagging, and astrometry.
Additionally, DR1 will include new data products that quantify the level of \HeI{}, shuttle glow, and auroral emission in each pixel measurement, enabling modeling of these features in the continuum model.
A more physically motivated continuum model, such as one that includes templates for the hydrogen and PAH features, could further improve our modeling by reducing the need for our data-driven uncertainty calibrations.
Finally, while we restricted this analysis to the Deep Fields, the all-sky data offer complementary advantages, such as adding sightlines closer to the Galactic Center with substantially larger $D$-factors.
Because the expected DM signal varies across the sky in a known way, whereas instrumental and atmospheric residuals are expected to have a different spatial dependence, an all-sky analysis could exploit the morphology of the signal to distinguish it from residual spectral structure that is not captured by our continuum model and is common to many sightlines (e.g., from instrumental calibration, zodiacal light modeling, or atmospheric emission). 
Realizing this potential will require modeling the brighter and more spatially and spectrally variable foregrounds at low ecliptic and Galactic latitudes, and we leave such an analysis to future work.

\begin{acknowledgments}
We thank Aurelio Amerio and
    Carolina Cuesta-L\'azaro
for helpful discussions;
    Marco Regis, Jorge Terol Calvo, and Marco Taoso
for a prompt exchange regarding their SPHEREx sensitivity projections;
    and Andrew Saydjari
for his collaboration in building the pixel-level database.
We especially thank the SPHEREx team for their extremely generous data release policy. 
CH is supported by the National Science Foundation Graduate Research Fellowship under Grant Number DGE-2234660.
The Flatiron Institute is a division of the Simons Foundation.
\end{acknowledgments}

\bibliography{main,axionlimits}{}
\bibliographystyle{aasjournalv7}

\appendix

\begin{figure*}[t]
    \centering
    \includegraphics[width=\linewidth]{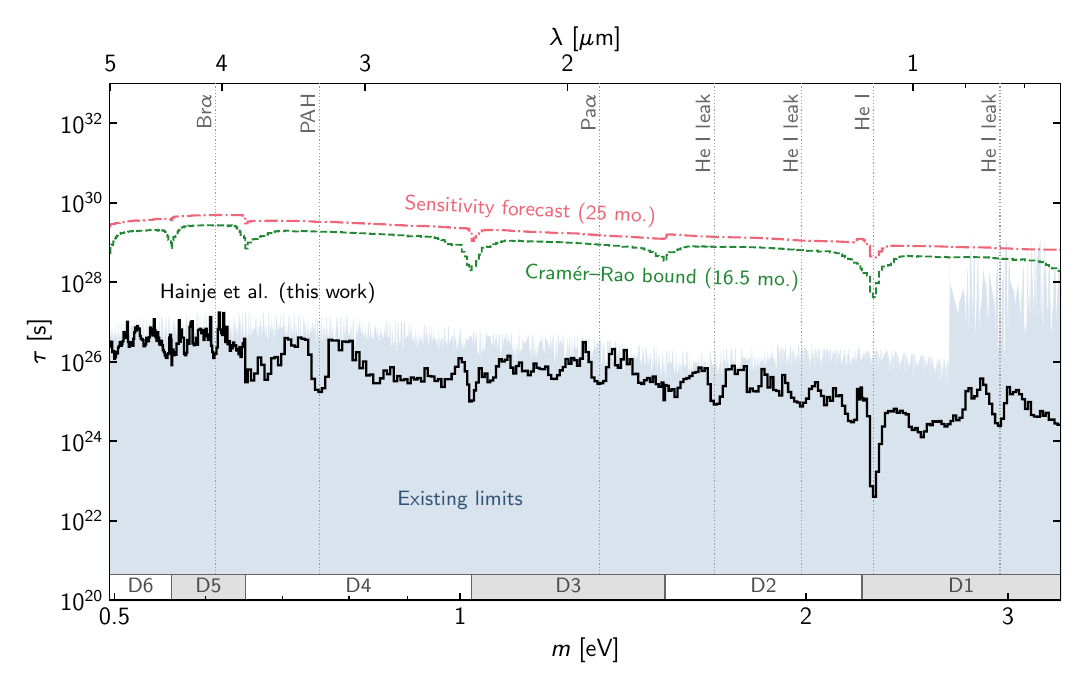}
    \caption{Model-independent 95\% power-constrained lower limits on the decay lifetime $\tau$ of the DM particle.}
    \label{fig:tau-constraint}
\end{figure*}

\section{Lifetime constraints}
\label{appx:tau}

Our upper limits on the emission rate can also be cast as lower limits on the lifetime $\tau = 1 / \Gamma$ of any DM particle that decays to two photons, independent of the axion model.
This is shown in Figure~\ref{fig:tau-constraint}.

\section{Profiled goodness of fit}
\label{appx:profiling}

We choose $\mathcal{B}$ to be a linear model of the data that is independently fit for each (spatial, detector) bin $b$, $\mathcal{B} = \mat{M}_b \, \vec{\theta}_b$, so we can analytically profile out the continuum parameters.
(Here $\mat{M}_b$ is the design matrix, with number of rows equal to the number of pixel measurements, and number of columns equal to the number of parameters.)
Let $\mat{C}_b = \text{diag}(\sigma_i^2)$ be the covariance matrix of the bin, let $\vec{d}_b$ be the zodiacal-light-subtracted intensities ($d_{b,i} = I_{\nu,i} - Z_i$), and let $\vec{a}_b$ be the axion intensity template ($a_{b,i} = I_\nu^\DM (\uvec\alpha_b ; j_i) / \Gamma$). 
Because the coupling $\gagg$ and the emission rate $\Gamma$ are related for a given mass $m_a$, it is convenient to proceed in terms of $\Gamma$.
Then the profiled goodness of fit is given by
\begin{align}
    \chi^2_{\rm p} (\Gamma \, | \, m_a)
    &=
    \min_{\vec\theta} \chi^2 (\Gamma, \vec\theta \, | \, m_a)
    \\ &=
    \sum_{b}
    \min_{\vec\theta_b} \,
    \big(
    \Gamma \, \vec{a}_b
    + \mat{M}_b \, \vec{\theta}_b
    - \vec{d}_b
    \big)^\top \, \mat{C}_b^{-1} \, \big(
    \Gamma \, \vec{a}_b
    + \mat{M}_b \, \vec{\theta}_b
    - \vec{d}_b
    \big)
    \\ &=
    \sum_{b}
    \underbrace{\vec{a}_b^\top \mat{P}_b \vec{a}_b}_{A_b} \,
    \Gamma^2
    - 2 \, 
    \underbrace{\vec{a}_b^\top \mat{P}_b \vec{d}_b}_{B_b} \,
    \Gamma
    + 
    \underbrace{\vec{d}_b^\top \mat{P}_b \vec{d}_b}_{C_b} ,
\end{align}
where we have introduced
\begin{equation}
    \mat{P}_b \equiv \mat{C}_b^{-1} - 
    \mat{C}_b^{-1}
    \mat{M}_b \,
    \big(
        \mat{M}_b^\top
        \mat{C}_b^{-1}
        \mat{M}_b
    \big)^{-1} \,
    \mat{M}_b^\top
    \mat{C}_b^{-1} .
\end{equation}
The profiled goodness of fit is quadratic in the emission rate $\Gamma$:
\begin{equation}
    \chi^2_{\rm p} (\Gamma \, | \, m_a) =
    \underbrace{\big( { \textstyle
        \sum_{b} A_b
    } \big)}_{A} \,
    \Gamma^2
    - 2 \, 
    \underbrace{\big( { \textstyle
        \sum_{b} B_b
    } \big)}_{B} \,
    \Gamma
    + 
    \underbrace{\big( { \textstyle
        \sum_{b} C_b
    } \big)}_{C} .
\end{equation}
The maximum-likelihood estimator is therefore $\hat\Gamma = B / A$, with uncertainty $\sigma_{\hat\Gamma} = 1/\sqrt{A}$.
Because the model is linear in $(\Gamma, \boldsymbol\theta_b)$ and if the noise is Gaussian with covariance $\mathsf{C}_b$, the Fisher information on $\Gamma$ after marginalizing over the continuum parameters is $\sum_b \boldsymbol{a}_b^\top \mathsf{P}_b \boldsymbol{a}_b = A$ and the Cram\'er--Rao bound for any unbiased estimator of $\Gamma$ is $\sigma_{\Gamma}^2 \geq 1/A$, which $\hat\Gamma$ saturates provided the true continuum lies in the span of $\mathsf{M}_b$.

Our specific choice of linear model is a fourth-order polynomial.
We choose as our design matrix $\mat{M}_{b,st} = [(\lambda_s - \lambda_a) / \Lambda]^t$, where $\lambda_a$ is the wavelength of the axion signal and $\Lambda$ is the half-width of the fitting window.
We use a fitting window with half-width equal to 5 times the local resolution; i.e., $\Lambda = 5 \, \lambda_a / R$.
Because two target axion masses spaced closer than $\lambda_a / R$ will share almost their entire fitting window, we in fact only fit the continuum model at the central wavelengths of the nominal spectral channels. 
Our search grid of axion masses $m_a$ is subdivided from the grid of spectral channels, so each mass uses the continuum model of its parent channel.

We tested a few variations on these modeling decisions.
We varied the window size ($\Lambda = 10 \, \lambda_a / R$) and the continuum model (sixth-order polynomial, spline with knot spacings at 2.5 and 3.5 times the local resolution $\lambda_a/R$) on a $0.5^\circ$-radius circular patch centered on the north ecliptic pole.
These experiments changed the final limits very little (by a maximum of a few percent), and so we have chosen to use the simplest model for the main analysis.

Because the emission rate cannot physically be negative, we in the end set our limits using $\Delta \chi^2 (\Gamma) = \chi_{\rm p}^2 (\Gamma) - \chi_{\rm p}^2 (\max(\hat\Gamma, 0))$, with the 95\% confidence upper limit set where $\Delta \chi^2 = 2.71$.
We can write this upper limit in terms of the quadratic coefficients:
\begin{equation}
    \Gamma_{95} = \frac{B + \sqrt{2.71 \, A + \min(B, 0)^2}}{A} .
\end{equation}

\section{Power-constrained limit}
\label{appx:pcl}

We also apply the power-constrained limit (PCL) of \citet{Cowan+2011-PCL}, in a similar fashion to \citet{JanishPinetti2025} and \citet{Pinetti2025}.
Fundamentally, PCL asks: given the best-fit continuum model, what upper limit would be attained with a $-1\sigma$ fluctuation?
The power-constrained limit is then given by the maximum of $\Gamma_{95}$ and this upper limit.

Under the null hypothesis, the pull $z = \hat\Gamma / \sigma_{\hat\Gamma}$ is normally distributed.
Also, notice that $\Gamma_{95} = (z + \sqrt{2.71 + \min(z,0)^2}) / \sqrt{A}$ is a monotonically increasing function of $z$.
Therefore, the $-1\sigma$ value of $\Gamma_{95}$ under the best-fit continuum model is simply $\Gamma_{95}$ evaluated at the $-1\sigma$ value of $z$, which is just $-1$.
The power-constrained limit is then
\begin{equation}
    \Gamma_{\rm PCL} = \max\big( \Gamma_{95}, (\sqrt{3.71} - 1)/\sqrt{A} \big) .
\end{equation}

\section{Data-driven error calibration}
\label{appx:error-calibration}

In an effort to be maximally conservative, we perform a number of data-driven corrections to inflate uncertainties where relevant.
These corrections were first described in \citet{Pinetti2025}.

\textbf{Correction I} ---
We first rescale the measurement uncertainties if they underestimate the empirical scatter in the spectrum.
For every bin $b$, we fit the continuum model as described previously.
Then, we compare the uncertainty $\sigma_i$ of each measurement to a sliding-window standard deviation over the wavelength-sorted, continuum-subtracted, zodiacal-light-subtracted intensities (using a window size of nine measurements).
We replace all measurement uncertainties with the maximum of the reported value and the sliding-window standard deviation, then repeat the continuum fit and the calculation of the quadratic coefficients $A_b, B_b$.

The next two corrections make use of the fact that,
under the null hypothesis ($\Gamma = 0$, Gaussian-distributed noise), the pull $z$, defined as the ratio of the maximum-likelihood estimator to its uncertainty $\hat\Gamma / \sigma_{\hat\Gamma}$, should be normally distributed, $z \sim \mathcal{N}(0, 1)$.
We therefore look for deviations from normality in the distributions of pulls.

\textbf{Correction II} ---
For each bin, we can look at the distribution of pulls across different target axion masses and rescale the pulls to normalize this scatter.
When treating target mass $m_a$, we look at the pulls of target masses with wavelengths $\lambda_a \pm n \, \lambda_a / R$ for $n \in \{ 2, 3, 4, 5, 6 \}$, where those neighbors land in the same detector.
(When near a detector edge, we expand the number of neighbors on one side up to $n = 11$.)
We then compute the robust width $S_b$, defined as 1.4826 times the median absolute deviation (MAD), of the pulls $z$ for these neighbors, and divide $A_b, B_b$ by a factor of $\max (1, S_b^{2})$.

\textbf{Correction III} ---
We perform the same rescaling to the total quadratic coefficients $A, B$.
For target mass $m_a$, we look at neighboring masses with wavelengths spaced by the spectral channel spacing $\lambda_a / R$, starting at two spacings away, up to ten neighbors as in Correction II.
Then we compute the robust width $S_{\rm III}$ of the total pulls and divide $A, B$ by $\max (1, S_{\rm III}^2)$.
Notice that because both quadratic coefficients are scaled by the same factor,
$\hat\Gamma = B / A$ is invariant to this scaling, and the uncertainty $\sigma_{\hat\Gamma} = 1/\sqrt{A}$ scales as $S_{\rm III}$.
We call this correction the \textit{spurious-signal correction}, because it inflates uncertainties according to the presence of spurious excursions over neighboring mass points.
This has the effect of requiring that an emission line be very strong to be detected.\\

In practice, Correction I has a modest effect: 49\% of measurements are inflated, and the median inflation is $1.20 \times$, raising the final uncertainties on $\gagg$ by about 4\%.\footnote{In a simulation of correct, white-noise uncertainties, the same procedure inflates 43\% of measurements by median $1.16 \times$, indicating that the reported SPHEREx uncertainties are generally close to the observed scatter.}
Correction II also has a rather modest effect, inflating the uncertainties on $\gagg$ by a median of $\sim 6\%$.
Correction III has a substantial effect, inflating the uncertainties on $\gagg$ by a median factor of 37.
This suggests the presence of correlated spectral residuals across the sky that are not adequately captured by our smooth continuum model, highlighting the importance of improved continuum modeling in future analyses.
Even after these corrections, the empirical distribution of $z$ has far heavier tails than a Gaussian (20.6\% of the tested masses have $|z| > 1.96$, compared to 5\% for a normal distribution). 
In particular, this means that $z$-scores (such as those in Table~\ref{tab:detections}) should not be treated as a number of sigma.

\end{document}